\documentclass[aps,prl,twocolumn,groupedaddress,amsmath,amsfonts,showpacs]{revtex4}

\usepackage[dvipdfmx]{graphicx}
\usepackage{bm} 
\usepackage{braket}
\usepackage[usenames]{color}
\usepackage{comment}

\begin{document}


\title{Electromagnetic and thermal responses of Z topological insulators and superconductors in odd spatial dimensions}


\author{Ken Shiozaki}
\author{Satoshi Fujimoto}
\affiliation{Department of Physics, Kyoto University, Kyoto 606-8502, Japan}




\date{\today}

\begin{abstract}
The relation between bulk topological invariants and experimentally observable physical quantities is
a fundamental property of topological insulators and superconductors.
In the case of chiral symmetric systems in odd spatial dimensions such as time-reversal invariant topological superconductors
and topological insulators with sublattice symmetry, 
this relation has not been well understood. 
We clarify that the winding number which characterizes the bulk Z non-triviality of these systems can appear in electromagnetic and thermal responses in a certain class of heterostructure systems.  
It is also found that the Z non-triviality can be detected in the bulk "chiral polarization", which is 
induced by magnetoelectric effects.
\end{abstract}



\maketitle

An important feature of topological insulators (TIs) and topological superconductors (TSCs) is that
topological invariants characterizing the bulk states
emerge as physical quantities probed by electromagnetic or thermal responses \cite{Review}.
For instance, the Chern number appears as
the quantized Hall conductivity in the quantum Hall effect state \cite{TKNN}, and 
the Z$_2$ invariant of a time-reversal invariant (TRI) TI in three dimensions can be detected in
axion electromagnetic responses \cite{QHZ}.
The correspondence between bulk topological invariants and electromagnetic (or thermal) responses naturally arises from
the existence of underlying low-energy effective topological field theories \cite{QHZ}.
For most classes of TIs and TSCs \cite{Schnyder,Kitaev2}, this correspondence has been well clarified so far.
However, for the case of TIs and TSCs characterized by Z invariants in odd spatial dimensions,
this point has not yet been fully understood. 
These classes include time-reversal symmetry broken (TRB) TIs with sublattice symmetry in one and three dimensions (class AIII),
TRI TSCs in three dimension (class DIII, e.g. $^3$He, Cu$_x$Bi$_2$Se$_3$ \cite{CuBiSe,Ando}, 
Li$_2$Pt$_3$B \cite{LiPtB}), 
and TRI TIs and TSC of spinless fermions in one dimension (class BDI, 
e.g. Su-Schrieffer-Heeger model \cite{SSH}, Kitaev Majorana chain model \cite{Kitaev}).
It is noted that all of these classes possess chiral symmetry (sublattice symmetry); i.e. the Hamiltonian $H$ satisfied the relation 
$\Gamma H \Gamma = -H$ with $\Gamma$ a unitary operator. 
This implies that if $|\psi\rangle$ is an eigen state of $H$ with an energy $E$,
then, $\Gamma |\psi\rangle$ is also an eigen state with an energy $-E$.
The chiral symmetry is indeed the origin of  
the bulk Z topological invariant referred to as the winding number.
The chiral symmetric topological insulator with the winding number $N$ possesses $N$ flavors of gapless Dirac (Majorana) fermions at the boundary, 
which are stable against disorder and interactions, as long as the chiral symmetry is preserved \cite{stability}.
To this date, however, it has not been fully elucidated how the winding number
can be detected in electromagnetic or thermal responses.  
For instance, in the case of three-dimensional (3D) class AIII TIs, 
low-energy effective theory is the axion field theory as in the case of TRI Z$_2$ TIs, the action of which is
given by \cite{QHZ,Ryu},
\begin{equation}
\begin{split}
S_{\rm axion} = \frac{e^2}{2 \pi \hbar c} \int dtd^3 x P_3 \bm{E} \cdot \bm{B}, 
\end{split}
\label{Axion}
\end{equation}
where 
\begin{equation}
\begin{split}
P_3 = \frac{1}{8 \pi^2} \int_{BZ} \mathrm{tr} \left[ \mathcal{A} d \mathcal{A} - \frac{2i}{3} \mathcal{A}^3 \right]
\end{split}
\label{MEP}
\end{equation}
is the magnetoelectric polarization expressed by the Chern-Simons 3-form 
with Berry connection $\mathcal{A}_{nm}(\bm{k}) = i \Braket{u_n(\bm{k}) | d u_m(\bm{k})}$ for occupied states $\ket{u_n(\bm{k})}$.
Because of chiral symmetry, $P_3$ takes only two values, i.e. $P_3 = \frac{N_3}{2}$ (mod 1) where $N_3$ is the integer-valued winding number~\cite{Ryu2}.
Thus, the above field theory captures only Z$_2$ part of the winding number, and fails to describe
the Z nontrivial character~\cite{Hosur}.
The same problem also occurs for class DIII TSCs, as previously noticed by Wang 
and his coworkers \cite{Wang}.
For this case, Wang et al. presented an argument based on an effective theory for surface Majorana fermions.
However, a general framework which relates the winding number to electromagnetic or thermal responses
is still lacking, and desired.
In this paper, we present two approaches for the solution of this issue.
One is based on the idea that the winding number can be detected in electromagnetic and thermal responses
of a certain class of heterostructure systems
(see Eqs.(\ref{Hall_cond})-(\ref{ME}) and (\ref{thermalHall})-(\ref{gme2}) below, which constitute main new results).
We clarify the condition for the heterostructure systems in which the Z non-trivial character of the bulk systems
can appear.
The other one is to introduce a novel bulk physical quantity which can be directly related to the winding number. 
This quantity is referred to as chiral charge polarization.
We show that for 3D class AIII TIs,
the chiral charge polarization is induced by an applied magnetic,
which is in analogy with topological magnetoelectric effect, and furthermore, the winding number appears in
its response function (see Eq.(\ref{P5B}) below).
\textit{Bulk winding number and magnetoelectric polarization in chiral symmetric TIs ---}
We, first, consider the approach based on heterostructure systems.
To explain our approach in a concrete way, we consider 3D chiral-symmetric (CS) TIs, i.e. a class AIII systems.
The following argument is straightforwardly extended to the case of class DIII TSCs.
A key idea is to consider 
a heterostructure system which consists of the 3D CS TI
and a chiral-symmetry-broken (CSB) trivial insulator with the Hamiltonian,
as depicted in FIG. 1(a).
Here, the trivial insulator means that $P_3=0$ in the bulk \cite{P3_0}. 
For instance, we can consider the CSB trivial insulator with inversion symmetry in the bulk which ensures $P_3=0$.
To deal with spatially varying heterostructure systems, we utilize
an adiabatic approach. 
That is, as long as there is a finite energy gap which separates the ground state and the first excited states,
the interface structure can be smoothly deformed to the slowly varying one.
In the slowly varying structure, the position operator $\hat z$ in the Hamiltonian can be treated as a parameter (adiabatic parameter) independent of 3D momentum $\bm{k}$, which parametrizes
the spatial inhomogeneity of the heterostructure.
Then, the magnetoelectric polarization $P_3(z)$ is 
constructed from the adiabatic Hamiltonian of the heterostructure system, $\tilde{H}(\bm{k},z)$. 
$\tilde{H}(\bm{k},z)$ interpolates between the bulk Hamiltonian of the CS TI, $H(\bm{k})$, 
and that of the CSB trivial insulator, $H_{\rm CSB}(\bm{k})$, when $z$ is varied; i.e.
$\tilde{H}(\bm{k},z)=H(\bm{k})$ when $z$ is a point in the bulk of the CS TI, and
 $\tilde{H}(\bm{k},z)=H_{\rm CSB}(\bm{k})$ when $z$ in the bulk of the CSB trivial insulator.
The adiabatic approach was exploited before to derive electromagnetic responses of 
the TRI Z$_2$ TIs from the axion field theory \cite{QHZ,TK}.
Our strategy is to extend the adiabatic argument for the Z$_2$ non-triviality to 
the Z nontrivial electromagnetic responses. 
We, first consider the quantum anomalous Hall effect.
We note that in the heterostructure junction system,
the anomalous Hall effect caused by surface Dirac fermions is obtained by integrating $z$-direction under a $z$-independent electromagnetic field, 
\begin{equation}
\begin{split}
&S_{surf}[A(t,x,y)] \\
&= \frac{e^2}{2 \pi \hbar c} \left( \int_{z_0,C}^{z_1} dz \frac{d P_3(z)}{d z} \right) \int dtd^2 x \epsilon_{\mu \nu \rho} A_{\mu} \partial_{\nu} A_{\rho}. 
\end{split}
\end{equation}
Here $z_0$ ($z_1$) is a point in the CS TI (CSB trivial insulator), and
$C$ is a path of $z$-integral.
Hence the Hall conductivity is given by,
\begin{equation}
\begin{split}
\sigma_{\rm H} = \frac{e^2}{2 \pi \hbar} \int_{z_0,C}^{z_1} dz \frac{d P_3(z)}{d z} = \frac{e^2}{h} \int_{z_0,C}^{z_1} d P_3(z).
\label{QAHE}
\end{split}
\end{equation}
There are two important remarks. 
First, although the magnetoelectric polarization $P_3$ is gauge-invariant only for mod $1$, the line integral of
a small difference of $P_3(z)$ is fully gauge-invariant. 
Second, $\sigma_{\rm H}$ is determined not only by the bulk magnetoelectric polarization at
the point $z_0$ and that at the point $z_1$, 
but also by a homologous equivalence class of the path $C$. 
This means that $\sigma_{\rm H}$ depends on the microscopic structure of the interface, and is not protected solely by the bulk topology. 
The concrete path $C$ is determined by 
the signs of the mass gaps of Dirac fermions on the surface of the CS TI. 
In our system, 
the mass gaps are generated by the chiral-symmetry breaking field induced by
the CSB trivial insulator at the surface \cite{P3CSB}. 
Here, we consider the case that the sign of the chiral-symmetry breaking field, and hence, 
that of the induced mass gaps are uniform on the interface
between  the CS TI and the CSB trivial insulator.
More precisely, the Hamiltonian $H_{\rm CSB}(\bm{k})$ satisfying this condition is generally expressed as
$H_{\rm CSB}(\bm{k}) = H_0(\bm{k}) + \alpha(\bm{k}) \Gamma$ where $\alpha(\bm{k})>0$ (or $<0$) for any $\bm{k}$,
and $\Gamma$ is the chiral symmetry operator mentioned before, and
$H_0(\bm{k})$ does not generate mass gaps of the surface Dirac fermions.
It is noted that $\Gamma$ itself plays the role of a chiral-symmetry breaking field.
Then the winding number $N_3$ for $H(\bm{k})$ and the magnetoelectric polarization $P_3(z)$ for the adiabatic Hamiltonian $\tilde{H}(\bm{k},z)$ satisfies the following relation,
\begin{eqnarray}
\int_{z_0,C}^{z_1} d P_3(z)=\pm \frac{N_3}{2}.
\label{N3P3}
\end{eqnarray}
This relation is one of our central new findings.
We present  a sketch of the proof of Eq.(\ref{N3P3}) below.
The details are given in the supplemental materials \cite{Sup}.
Since the magnetoelectric polarization of the CS TI with the winding number $N_3$ is $P_3 = \frac{N_3}{2} \ \mathrm{mod} \ 1$, 
the value of $P_3(z=z_0)$ is fixed by chiral symmetry. 
Also, as mentioned above,
the value of $P_3(z = z_1)$ in the CSB trivial insulator is fixed to be zero. 
Due to these fixed boundary values, $\int_{z_0,C}^{z_1} d P_3(z)$ is adiabatically invariant, which means that this quantity is not changed unless the energy gap between the ground state and first excited state closes. 
Thus, we can deform $\tilde H$ to a flat band system : $\tilde H^2 = 1$. 
Note that $H_{\rm CSB}(\bm{k})$ mentioned above is deformed to $H_{\rm CSB}=\pm\Gamma$
without closing the energy gap.
Then, the adiabatically equivalent class of flat band Hamiltonian $\tilde H$ 
is given by
\begin{eqnarray}
\tilde H (\bm{k},\theta) = \cos \theta Q(\bm{k}) \pm \sin \theta \Gamma ,
\label{flatH}
\end{eqnarray}
where $Q(\bm{k}) = 1-2 P(\bm{k})$, and $P(\bm{k})$ is a projection to the occupied bands of $H(\bm{k})$, 
and $\theta$ monotonically changes from $\theta = 0$ to $\theta = \pi/2$, as $z$ changes from $z_0$ to $z_1$. 
It is straightforward to show 
$\int_{z_0,C}^{z_1} d P_3(z) = \int_{0}^{\frac{\pi}{2}} d P_3(\theta) = \pm \frac{N_3}{2}$ \cite{Sup}, where $P_3(\theta)$ is the magnetoelectric polarization of $\tilde H(\bm{k},\theta)$ defined by (\ref{MEP}). 

\begin{figure}[!]
 \begin{center}
  \includegraphics[width=\linewidth, trim=0cm 0cm 0cm 0cm]{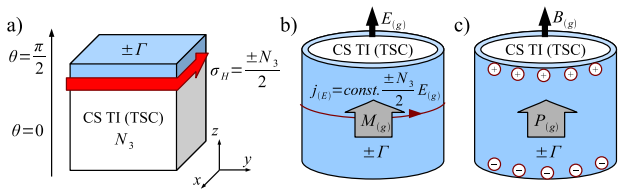}
 \end{center}
 \caption{Heterostructure composed of 
 a CS TI (or a CS TSC) 
 and trivial insulators (or superconductors) with Hamiltonian $H = \pm \Gamma$.
 In (b) and (c), we assume CS TI (TSC) is coated by a CSB trivial insulator (SC) so that the interface structure has a finite energy gap anywhere.
 }
 \label{chiralTI}
\end{figure}

Using Eqs.(\ref{N3P3}) and (\ref{QAHE}) together, we can readily obtain the remarkable result that
the winding number $N_3$ can appear in
the quantized Hall conductivity for
the heterostructure system depicted in FIG.1(a),
\begin{equation}
\begin{split}
\sigma_{\rm H} = \pm \frac{e^2}{2h}N_3.
\label{Hall_cond}
\end{split}
\end{equation}
Hence, the Z non-triviality of CS TIs can be detected experimentally in this electromagnetic response.

We can also apply
the formula (\ref{N3P3}) to the investigation on topological magnetoelectric effects which are
characterized not by the Z$_2$ invariant, but by the Z invariant $N_3$. 
Let us consider the heterostructure system depicted in FIG. 1(b) and (c), which
 consists of a cylindrical CS TI with its surface coated by a CSB trivial insulators.
From Eq. (\ref{N3P3}), the magnetoelectric polarization of the CS TI coated by the CSB trivial insulator is given by $P_3 = P_3(z=z_1) - \int_{z_0,C}^{z_1} d P_3(z) = \mp \frac{N_3}{2} $, 
which leads the magnetoelectric effect, $\bm{P}= - \frac{e^2}{hc} P_3 \bm{B}$ and $\bm{M}=-\frac{e^2}{hc}P_3\bm{E}$, \cite{QHZ} i.e.,  
\begin{eqnarray}
\bm{P}=\pm\frac{e^2}{2hc}N_3\bm{B},
\label{PB}
\end{eqnarray}
\begin{eqnarray}
\bm{M}=\pm\frac{e^2}{2hc}N_3\bm{E}.
\label{ME}
\end{eqnarray}
Here, $\bm{E}$ and $\bm{B}$ are an electric field and a magnetic field applied parallel to the axis of the cylinder.
The winding number successfully appears in the above magnetoelectric responses.
It is noted that if the system is extended without open boundaries and possesses translational symmetry, 
magnetoelectric polarization (\ref{MEP}) is gauge-dependent under large gauge transformation so that $P_3 \mapsto  P_3 + n$ where $n$ is integer. 
$P_3 = \mp \frac{N_3}{2}$ in Eqs. (\ref{PB}) and (\ref{ME}) implies that the particular choice of the configuration of the heterostructure as depicted in FIG.1(b) and (c) corresponds to the particular choice of the gauge that can extract the winding number of the TI in the heterostructure system.


\textit{Case of TRI TSCs ---}
The above argument is also applicable to class DIII TRI TSCs in three dimensions. 
In the case of TSCs with spin-triplet pairing, since both charge and spin are not 
conserved,
it is difficult to detect the topological character in electromagnetic responses. 
However, instead, thermal responses can be a good probe for the topological nontriviality, because
surface Majorana fermions still preserve energy. 
An effective low energy theory for the thermal responses of TSCs is the gravitational axion field theory described by the action \cite{Ryu, Nomura, note}, 
\begin{equation}
\begin{split}
S = \frac{\pi k_B^2 T^2}{12 \hbar v} \int dt d^3 x P_3(x) \bm{E}_g \cdot \bm{B}_g
\end{split}
\label{TSC-axion}
\end{equation}
where $\bm{E}_g$ is a gravitoelectric field which play the same role as
temperature gradient $- \bm{\nabla} T/T$, and $\bm{B}_g$ is a gravitomagnetic field, 
which is in analogy with a magnetic field of electromagnetism,
and $v$ is the fermi velocity.  
Because of time-reversal symmetry, $P_3$ in the above action (\ref{TSC-axion}), takes only two values, i.e.
$0$ or $1/2$, implying the Z$_2$ non-triviality, and hence Eq.(\ref{TSC-axion}) is an incomplete description for the Z nontrivial TSCs.
However, as in the case of class AIII TIs discussed above,
the winding number $N_3$ can be detected as thermal responses in a certain class of heterostructure system.
In the case of class AIII TIs, an important role is played by the chiral-symmetry-breaking field $\Gamma$. 
Similarly, also in the case of TRI TSCs, the winding number appears in the heterostructure system composed of 
a TRI TSC and a trivial phase with broken chiral-symmetry.
For a $p$-wave TSC which is realized in $^3$He, Cu$_x$Bi$_2$Se$_3$  and Li$_2$Pt$_3$B, 
the chiral-symmetry-breaking field is nothing but an s-wave pairing gap with broken time reversal symmetry.
This is easily seen from the fact that 
the s-wave pairing term of the Hamiltonian is expressed as 
${\rm Re}\Delta_s\tau_y\sigma_y+{\rm Im}\Delta_s\Gamma$
where $\Delta_s$ is the s-wave gap, and $\tau_\mu$ ($\sigma_\mu$) is
the Pauli matrix for particle-hole (spin) space, and the chiral symmetry operator $\Gamma$ is expressed as 
$\Gamma=\tau_x\sigma_y$.
When the imaginary part of $\Delta_s$ is nonzero, this term breaks chiral-symmetry.
Thus, Eq.(\ref{N3P3}) is applicable for the heterostructure system composed of a TRI TSC and a trivial $s$-wave SC with broken time-reversal symmetry,
as long as the real part of the s-wave gap does not yield gap-closing.
For the system depicted in FIGs.1(a), (b) and (c)
the quantum anomalous thermal Hall effect and the topological gravitomagetoelectric effects associated with
the winding number are realized.
Combining the gravitational axion field theory (\ref{TSC-axion}) and the relation (\ref{N3P3}),
we obtain the quantum anomalous thermal Hall conductivity, 
\begin{eqnarray}
\kappa_{xy}=\frac{\pi^2 k_{\rm B}^2T}{12h}N_3,
\label{thermalHall}
\end{eqnarray}
realized for the system shown in FIG.1(a).
This result essentially coincides with that obtained by Wang et al. from the argument based on
surface Majorana fermions \cite{Wang}.
We can also obtain the gravitomagnetoelectric effects,
\begin{eqnarray}
\bm{P}_g= \pm\frac{\pi^2k_{\rm B}^2T^2}{12hv}N_3\bm{B}_g,
\label{gme1}
\end{eqnarray}
\begin{eqnarray}
\bm{M}_g= \pm \frac{\pi^2k_{\rm B}^2T^2}{12hv}N_3\bm{E}_g,
\label{gme2}
\end{eqnarray}
realized for the system shown in FIG.1(b) and (c).
Eq.(\ref{gme1}) implies that circulating energy current flows surrounding 
the axis of the cylinder
induces the energy (or thermal) polarization, resulting in nonzero temperature gradient along the axis.
The winding number explicitly appears in this thermal response.

\textit{Chiral polarization and the winding number---}
Hitherto, we have explored the Z topological responses in heterostructure junction systems in which
the winding number successfully emerges as the quantum Hall effect and the topological magnetoelectric effect. 
However, it is still desirable to establish a direct connection between the winding number and the bulk physical quantities,
as in the case of the quantum Hall effect in a two-dimensional electron gas and Z$_2$ TIs. 
We pursue this possibility here. 
For this purpose, we introduce the chiral polarization defined by,
\begin{eqnarray}
\bm{P}^5 = \frac{e}{V_c} \sum_{n \in \mbox{occupied}}\langle w_{n}|\hat{\bm{X}}^5|w_{n}\rangle,
\label{chiralp}
\end{eqnarray}
where $|w_{n}\rangle$ is the Wannier function, $V_c$ is the unit cell volume,  
and $\hat{\bm{X}}^5$ is the projected chiral position operator defined by
$\hat{ X}^5_{\mu} = P \Gamma \hat r_{\mu} P$ with 
$\hat{\bm{r}}$ a position operator and
$P$ the projection to the occupies states.
Generally, to construct the Wannier function 
localized exponentially in real space, we need the absence of gauge obstruction of the Bloch wave function, 
i.e., vanishing of Chern number $C_{ij}/(2 \pi i) = \int_{BZ} d^3k/(2 \pi)^3 \mathrm{tr} \mathcal{F}_{ij} = 0$ \cite{Marzari}. 
In chiral symmetric systems, the Chern numbers $C_{ij}$ are zeroes \cite{chiral_Chern}, and  
hence the exponentially localized Wannier functions are always well defined.
Eq. (\ref{chiralp}) is similar to charge polarization, but an important difference is that the chiral symmetry operator $\Gamma$
is inserted in (\ref{chiralp}). 
For the class AIII TIs and the class BDI TIs with two sub-lattice structures, 
$\bm{P}^5$ represents a difference of charge polarization between two sub-lattices.
It is noted that in contrast to charge polarization which depends on the choice of gauge,
$\bm{P}^5$ is gauge-invariant, 
since the gauge ambiguity cancels out between the two sub-lattice contributions \cite{Sup}. 
As will be shown below, $\bm{P}^5$ is a key bulk quantity which can be related to the winding number.
Actually, in the case of one-dimensional (1D) systems, $\bm{P}^5$ is expressed by the 1D winding number
$N_1$ as \cite{Sup},
\begin{eqnarray}
P^5= -\frac{N_1 e}{2}.
\label{P5N1}
\end{eqnarray}
For instance, for the 1D BDI class TIs such as the Su-Schrieffer-Heeger model of polyacetylene,
Eq.(\ref{P5N1}) represents fractional charges which appear at open edges of the system.
Eq.(\ref{P5N1}) is derived from non-trivial algebraic properties satisfied by
$\hat{\bm{X}}^5$, which can be regarded as a generalization of the commutation relation of the projected position operator $P\hat{\bm{r}}P$ \cite{Sup}.  
In the 3D case, this algebra also yields an interesting result that
the winding number $N_3$ is expressed by
the Nambu three bracket of $\hat{\bm{X}}^5$\cite{Nambu1,Sup}, 
which recently attracts much attentions in connection with the density algebra in 3D TIs \cite{neu,bern}.
However, we have not yet succeeded to relate the Nambu bracket to any physical quantities in condensed matter systems.
Thus, we here take a different approach for the 3D case.
In fact, in the case of 3D AIII TIs, 
on the assumption that the occupied and unoccupied Wannier states satisfy the chiral symmetry $\Ket{w_{\bar n}} = \Gamma \Ket{w_{n}}$ ($\bar n \in \text{unoccupied}, n \in \text{occupied}$ ),
a more remarkable and useful relation between $\bm{P}^5$ and the winding number $N_3$  can be derived;
$\bm{P}^5$ can be induced by an applied magnetic field, in analogy with the topological magnetoelectric effect, and furthermore, $N_3$ appears in the response function.
From the first-order perturbative calculation with respect to a magnetic field,
we obtain \cite{Sup},
\begin{eqnarray}
\bm{P}^5=-\frac{e^2}{2hc}N_3\bm{B}.
\label{P5B}
\end{eqnarray} 
Thus, the winding number can be detected as the chiral polarization induced by a magnetic field.
This is another main result of this paper. 

It is expected that an analogous effect may be realized in 3D TRI TSCs.
In the case of TSCs, to explore topological characters, we need to consider thermal responses, instead of electromagnetic ones. However, we have not yet succeeded to obtain thermal analogue of Eq.(\ref{P5B}).
Furthermore, it is highly non-trivial what $\bm{P}^5$ means for the case of superconductors.
These are important open issues which should be addressed in the near future.

\textit{Conclusion --- }
We have clarified that the Z non-triviality of 3D TRI TSCs and TIs with sub-lattice symmetry
can appear in electromagnetic and thermal responses of heterostructure systems
which consist of the TSCs or TIs and CSB trivial s-wave superconductors or band insulators.
We have also established the relation between the bulk winding number and the bulk chiral polarization,
which may be utilized for experimental detection of the Z non-triviality.

The authors thank M. Sigrist, T. Neupert, and A. Shitade for fruitful discussions. 
This work is supported by the Grant-in-Aids for Scientific
Research from MEXT of Japan (Grants No. 23102714 and No. 23540406), 
and the Global COE Program 
``The Next Generation of Physics, Spun from Universality and Emergence.''




\begin{thebibliography}{99}
\bibitem{Review} M. Z. Hasan and C. L. Kane, Rev. Mod. Phys. {\bf 82}, 3045 (2010); 
X. L. Qi and S. C. Zhang, Rev. Mod. Phys. {\bf 83}, 1057 (2011).
\bibitem{TKNN} D. J. Thouless, M. Kohmoto, M. P. Nightingale, and M. den Nijs,
Phys. Rev. Lett. {\bf 49}, 405 (1982).
\bibitem{QHZ} X.-L. Qi, T. L. Hughes, and S.-C. Zhang, Phys. Rev. B {\bf 78}, 195424 (2008). 
\bibitem{Schnyder} A. P. Schnyder, S. Ryu, A. Furusaki, and A. W. W. Ludwig, Phys. Rev. B {\bf 78}, 195125 (2008). 
\bibitem{Kitaev2} A. Kitaev, AIP Conf. Proc. {\bf 1134}, 22 (2009).
\bibitem{CuBiSe} Y. S. Hor, A. J. Williams, J. G. Checkelsky, P. Roushan, J. Seo, Q. Xu, H. W. Zandbergen, A. Yazdani, N. P. Ong, and R. J. Cava, Phys. Rev. Lett. {\bf 104}, 057001 (2010)
\bibitem{Ando} S. Sasaki, M. Kriener, K. Segawa, K. Yada, Y. Tanaka, M. Sato, and Y. Ando, Phys. Rev. Lett.
{\bf 107}, 217001 (2011).
\bibitem{LiPtB} H. Q. Yuan, D. F. Agterberg, N. Hayashi, P. Badica, D. Vandervelde, K. Togano, M. Sigrist, and
M. B. Salamon, Phys. Rev. Lett. {\bf 97}, 017006 (2006).
\bibitem{SSH} W. P. Su, J. R. Schrieffer, and A. J. Heeger, Phys. Rev. B{\bf 22}, 2099 (1980).
\bibitem{Kitaev} A. Kitaev, Physics-Uspekhi {\bf 44}, 131 (2001).
\bibitem{stability}
As pointed out by Schnyder et al. \cite{Schnyder}, the $N$ flavors of the gapless Dirac (Majorana) fermions on the boundary of the chiral symmetric topological insulator are 
robust against arbitrary static perturbations with the chiral symmetry. 
Since static perturbations preserving chiral symmetry are described by gauge fields (Eq.(40) in ref.\cite{Schnyder}),
their effect is just to shift the Dirac points, and hence gapless features of Dirac (Majorana) fermions are protected. 
(Also see ref. \cite{Bernard}.)
\bibitem{Bernard} A. LeClair and D. Bernard, J. Phys. A{\bf 45}, 435203 (2012).
\bibitem{Ryu} S. Ryu, J. E. Moore, and A. W. W. Ludwig, Phys. Rev. B{\bf 85}, 045104 (2012).
\bibitem{Ryu2} S. Ryu, A. P Schnyder, A. Furusaki and A. W W Ludwig, New J. Phys.{\bf 12}, 065010 (2010). 
\bibitem{Hosur} P. Hosur, S. Ryu, and A. Vishwanath, Phys. Rev. B {\bf 81}, 045120 (2010).
\bibitem{Wang} Z. Wang, X. L. Qi, and S. C. Zhang, Phys. Rev. B{\bf 84}, 014527 (2011).
\bibitem{P3_0} Generally, it is possible that $P_3\neq 0$
even for a trivial insulator when both of time-reversal symmetry and inversion symmetry are broken. 
We do not consider such a specific case in our scenario.
\bibitem{TK} J. C. Y. Teo and C. L. Kane, Phys. Rev. B {\bf 82}, 115120 (2010).
\bibitem{P3CSB} $H_{\rm CSB}$ may contain terms proportional to
an identity matrix in the sublattice space, which break chiral symmetry, but do not generate mass gaps of the surface Dirac fermions. 
Since such terms merely shift the chemical potential, and do not affect our argument, we neglect them.
\bibitem{Sup} Supplemental material.
\bibitem{Nomura} K. Nomura, S. Ryu, A. Furusaki, and N. Nagaosa, Phys. Rev. Lett. {\bf 108}, 026802 (2012).
\bibitem{note} 
The derivation of the gravitoelectromagnetic axion action (\ref{TSC-axion}) by Nomura et al.\cite{Nomura} is based on the anomalous thermal Hall effect on the surface of TSC with a finite energy gap. 
On the other hand, Stone pointed out that the uniform gravitational field can not produce a non-trivial Riemann curvature in gravitational instanton term which is the source of the anomalous thermal Hall current \cite{Stone}.
Thus, it is not yet clear how the axion action (\ref{TSC-axion}) can be related to the gravitational anomaly in (3+1) dimensions discussed in refs. \cite{Ryu, Wang}.  
However, a recent careful analysis \cite{Niu,Sumiyoshi} which includes energy magnetization corrections
 revealed that bulk thermal Hall currents in two-dimensional gapped systems can be induced by
temperature gradient, supporting the argument in ref. \cite{Nomura}.
Thus, we believe that, as done in the main text, 
it is legitimate to discuss the thermal Hall effect and gravitomagnetoelectric effects as bulk effects in gapped systems
on the basis of Eq. (\ref{TSC-axion}), though the clarification of the topological origin of (\ref{TSC-axion}) needs further investigations.
Also, we note that Hidaka et al. \cite{Hidaka} proposed another origin of the gravitational axion action, which is based on the Nier-Yahn term \cite{Nieh}.
\bibitem{Stone} M. Stone, Phys. Rev. B {\bf 85}, 184503 (2012).
\bibitem{Niu}  T. Qin, Q. Niu, and J. Shi, Phys. Rev. Lett. 107, 236601 (2011).
\bibitem{Sumiyoshi} H. Sumiyoshi and S. Fujimoto, arXiv:1211.5419.
\bibitem{Hidaka} Y. Hidaka, Y. Hirono, T. Kimura, Y. Minami, arXiv:1206.0734. 
\bibitem{Nieh} H. T. Nieh and M. L. Yan, J. Math. Phys. {\bf 23}, 373 (1982). 
\bibitem{Marzari} C. Brouder, G. Panati, M. Calandra, C. Mourougane, and N. Marzari, Phys. Rev. Lett. {\bf 98}, 046402 (2007).
\bibitem{chiral_Chern}From the chiral symmetry, we can choose the unoccupied states $\ket{\bar n} = \Gamma \ket{n}$, 
then the Chern number of the {\it unoccupied} state, $\bar C_{ij}$ is identical with one of the occupied state $C_{ij}$ : $\bar C_{ij} = C_{ij}$ because 
$\mathrm{tr} \mathcal{F}^{(\bar n)}_{ij} = i \braket{\partial_i \bar n |\partial_j \bar n} - (i\leftrightarrow j) = i \braket{\partial_i n |\partial_j n} - (i\leftrightarrow j) = \mathrm{tr} \mathcal{F}^{(n)}_{ij}$. 
On the other hand, the sum of the Chern number for all bands is zero : $C_{ij} + \bar C_{ij}=0$ because 
$\sum_{n} \mathrm{tr} \mathcal{F}^{(n)} = \sum_{n} i \braket{\partial_i n|\partial_j n} -(i\leftrightarrow j) 
= -i \sum_{nm} \left[ \braket{n|\partial_i m}\braket{m|\partial_j n} - \braket{n|\partial_j m}\braket{m|\partial_i n} \right] = 0$.
Hence $C_{ij} = 0$. 
\bibitem{Nambu1} Y. Nambu, Phys. Rev. D{\bf 7}, 2405 (1973).
\bibitem{neu}  T. Neupert, L. Santos, S. Ryu, C. Chamon, and C. Mudry, Phys. Rev. B {\bf 86}, 035125 (2012).
\bibitem{bern} B. Estienne, N. Regnault, and B. A. Bernevig, arXiv:1202.5543.

\end{thebibliography}

\begin{thebibliography}{99}
\bibitem{Hatsugai} See for example, Y. Hatsugai, New J. Phys. {\bf 12}, 065004 (2010).
\bibitem{Schnyder} A. P. Schnyder, S. Ryu, A. Furusaki, and A. W. W. Ludwig, Phys. Rev. B {\bf 78}, 195125 (2008). 
\bibitem{Nambu} Y. Nambu, Phys. Rev. D{\bf 7}, 2405 (1973).
\bibitem{Kita} T. Kita and M. Arai, J. Phys. Soc. Jpn. {\bf 74}, 2813 (2005). 
\bibitem{Luttinger} J. M. Luttinger, Phys. Rev. {\bf 84}, 814 (1951). 
\end{thebibliography}

\renewcommand\theequation{S\arabic{equation}}
\setcounter{equation}{0}

\onecolumngrid

\section{Supplemental Material}

\subsection{The derivation of Eq. (5)}
In this section, we present a detailed derivation of Eq. (5). 
We calculate the continuous change of the magnetoelectric polarization $P_3(z)$ between the chiral-symmetric topological insulator $H(\bm{k})$ with winding number $N_3$ 
and chiral-symmetry-broken trivial insulator $H_{\text CSB}(\bm{k}) = H_0(\bm{k}) + \alpha(\bm{k}) \Gamma$ with a trivial magnetoelectric polarization $P_3 = 0$. 
As noted in the main text, we assume that $\alpha(\bm{k}) >0$ (or $<0$) for any $\bm{k}$, and $H_0(\bm{k})$ does not generate mass gaps. 
Thus all of the signs of the mass gaps of the surface Dirac fermions are determined by the sign of $\alpha(\bm{k})$. 
Since the value of $\int_{z_0,C}^{z_1} d P_3(z)$ is adiabatically protected against smooth deformation of surface structure which does not close the energy gap, it is sufficient for our purpose to consider a flat band system, and assume, without loss of generality, that the spatial inhomogeneity of the heterostructure system is sufficiently slow, allowing 
the semiclassical treatment of the spatially varying parameter.
Then, the Hamiltonian of our system is 
expressed in the form of Eq. (6) in the main text, 
\begin{equation}
\begin{split}
\tilde H(\bm{k},\theta) = \cos \theta Q(\bm{k}) \pm \sin \theta \Gamma, 
\end{split}
\label{tildeH}
\end{equation}
where $Q(\bm{k})$ is the "Q-function" defined by $Q(\bm{k}) = 1 - 2 P(\bm{k})$ with $P(\bm{k})$ the projection to occupied bands of $H(\bm{k})$. 
The continuous change of the magnetoelectric polarization $P_3(z)$ is given by,
\begin{equation}
\begin{split}
\int_{z_0,C}^{z_1} d P_3(z) = \int_{0}^{\frac{\pi}{2}} d\theta \frac{d P_3(\theta)}{d \theta} = \frac{1}{8 \pi^2} \int_{\theta = 0}^{\theta = \frac{\pi}{2}} \int_{BZ} \mathrm{tr} \mathcal{F}^2(\bm{k},\theta), 
\end{split}
\label{S2}
\end{equation}
where $P_3(\theta)$ is the magnetoelectric polarization of $\tilde H(\bm{k},\theta)$, $\mathcal{F} = d \mathcal{A} + \mathcal{A} \wedge \mathcal{A}$ is the Berry curvature, 
and $\mathcal{A}(\bm{k},\theta) = i \Braket{\tilde u(\bm{k},\theta) | d \tilde u(\bm{k},\theta)}$ is the Berry connection for the occupied states $\Ket{\tilde u(\bm{k},\theta)}$ of the semiclassical Hamiltonian $\tilde H(\bm{k},\theta)$.

Note that the Q-function of $\tilde H(\bm{k},\theta)$ is equivalent to $\tilde H(\bm{k},\theta)$ itself as shown below. 
The occupied states $\Ket{\tilde u (\bm{k},\theta)}$ of $\tilde H(\bm{k},\theta)$ are given by 
\begin{equation}
\begin{split}
\Ket{\tilde u (\bm{k},\theta)} = \cos \frac{\theta}{2} \Ket{u(\bm{k})} \mp \sin \frac{\theta}{2} \Gamma \Ket{u(\bm{k})}
\end{split}
\end{equation} 
with $\Ket{u(\bm{k})}$ the occupied states of $Q(\bm{k})$, $Q(\bm{k}) \Ket{u(\bm{k})} = - \Ket{u(\bm{k})}$. 
Then the Q-function of $\tilde H(\bm{k},\theta)$ is 
\begin{equation}
\begin{split}
\tilde Q(\bm{k},\theta) 
&= 1 - 2 \tilde P(\bm{k},\theta) \\
&= 1 - 2 \sum_{u \in {\text occupied}} \left( \Ket{\tilde u (\bm{k},\theta)} \Bra{\tilde u (\bm{k},\theta)} \right) \\
&= \cos \theta \sum_{u \in {\text occupied}} \Big( \Gamma \Ket{u(\bm{k})} \Bra{u(\bm{k})} \Gamma - \Ket{u(\bm{k})} \Bra{u(\bm{k})} \Big) \pm \sin \theta \sum_{u \in {\text occupied}} \Big( \Gamma \Ket{u(\bm{k})} \Bra{u(\bm{k})} + \Ket{u(\bm{k})} \Bra{u(\bm{k})} \Gamma \Big) \\
&= \cos \theta Q(\bm{k}) \pm \sin \theta \Gamma .\\
\end{split}
\label{tildeQ}
\end{equation}
Here, we have used the relation $\Gamma P(\bm{k}) + P(\bm{k}) \Gamma = \Gamma$ obtained from the chiral symmetry. 

We now calculated the right-hand side of (\ref{S2}).
Generally, the Chern form $\mathrm{tr} \mathcal{F}^{n}$ can be written in terms of the gauge invariant projection \cite{Hatsugai}.
In the case of $n = 2$, 
\begin{equation}
\begin{split}
\mathrm{tr} \mathcal{F}^2(\bm{k},\theta) = - \mathrm{tr} \left[ \tilde P(\bm{k},\theta) \left( d \tilde P(\bm{k},\theta) \right)^2 \right]^2, 
\end{split}
\end{equation}
or equivalently, with the use of $d \tilde P \tilde P + \tilde P d \tilde P = d \tilde P$ and $d \tilde Q = - 2 d \tilde P$, 
we have,
\begin{equation}
\begin{split}
\mathrm{tr} \mathcal{F}^2(\bm{k},\theta) = \frac{1}{2^5} \mathrm{tr} \left[ \tilde Q(\bm{k},\theta) \left( d \tilde Q(\bm{k},\theta) \right)^4 \right].  
\end{split}
\label{S6}
\end{equation} 
Dividing the external differential $d$ into $(d_{\bm{k}}, d_{\theta}) = (d_{k_x}, d_{ky}, d_{k_z}, d_{\theta})$, 
and using Eq.(\ref{tildeQ}),
we rewrite the right-hand side of Eq.(\ref{S6}) as,
\begin{equation}
\begin{split}
\mathrm{tr} \left[ \tilde Q \left( d \tilde Q \right)^4 \right] 
&= \mathrm{tr} \left[ \tilde Q \left\{ \left( d_{\bm{k}} \tilde Q \right)^3 \wedge d_{\theta} \tilde Q + \left( d_{\bm{k}} \tilde Q \right)^2 \wedge d_{\theta} \tilde Q \wedge d_{\bm{k}} \tilde Q + d_{\bm{k}} \tilde Q \wedge d_{\theta} \tilde Q \wedge \left( d_{\bm{k}} \tilde Q \right)^2 + d_{\theta} \tilde Q \wedge \left( d_{\bm{k}} \tilde Q \right)^3 \right\} \right] \\
&= 4 \ \mathrm{tr} \left[ \tilde Q \left( d_{\bm{k}} \tilde Q \right)^3 \wedge d_{\theta} \tilde Q \right] \\
&= 4 \ \mathrm{tr} \left[ \Big( \cos \theta Q(\bm{k}) \pm \sin \theta \Gamma \Big) \Big( \cos \theta d_{\bm{k}} Q(\bm{k}) \Big)^3 \wedge \Big( -\sin \theta Q(\bm{k}) \pm \cos \theta \Gamma \Big) d \theta \right] \\
&= \pm 4 \ \mathrm{tr} \left[ \Gamma Q(\bm{k}) \Big( d_{\bm{k}} Q(\bm{k}) \Big)^3 \right] \wedge \cos^3 \theta d \theta. 
\end{split}
\label{trQ}
\end{equation}
Here, at the second line of eq. (\ref{trQ}), we used a cyclicity of the trace and $d_{\bm{k}} \tilde Q \tilde Q = - \tilde Q d_{\bm{k}} \tilde Q $, 
and at the forth line, we used $\Gamma Q(\bm{k}) + Q(\bm{k}) \Gamma = 0$ followed from the chiral symmetry. 
For the basis in which $\Gamma$ is represented as $\Gamma = \begin{pmatrix}
1 & 0 \\
0 & -1
\end{pmatrix}$, 
$Q(\bm{k})$ is expressed as $Q(\bm{k}) = \begin{pmatrix}
0 & q(\bm{k}) \\
q^{\dag}(\bm{k}) & 0
\end{pmatrix}$ with a unitary matrix $q(\bm{k})$.
 Then, from Eqs. (\ref{S6}) and (\ref{trQ}), we have 
\begin{equation}
\begin{split}
\frac{1}{8 \pi^2} \mathrm{tr} \mathcal{F}^2(\bm{k},\theta) 
= \pm \frac{1}{32 \pi^2} \mathrm{tr} \left[ q^{\dag}(\bm{k}) d_{\bm{k}} q(\bm{k}) \right]^3 \wedge \cos^3 \theta d \theta. 
\end{split}
\end{equation}
Hence the change of the magnetoelectric polarization (\ref{S2}) is given by 
\begin{equation}
\begin{split}
\int_{z_0,C}^{z_1} d P_3(z) 
&= \pm \frac{1}{32 \pi^2} \int_{BZ} \mathrm{tr} \left[ q^{\dag}(\bm{k}) d_{\bm{k}} q(\bm{k}) \right]^3 \int_{\theta = 0}^{\theta = \frac{\pi}{2}} \cos^3 \theta d \theta \\
&= \pm \frac{1}{48 \pi^2} \int_{BZ} \mathrm{tr} \left[ q^{\dag}(\bm{k}) d_{\bm{k}} q(\bm{k}) \right]^3 \\
&= \pm \frac{1}{2} N_3, \\
\end{split}
\end{equation}
where $N_3 = \frac{1}{24 \pi^2} \int_{BZ} \mathrm{tr} \left[ q^{\dag}(\bm{k}) d_{\bm{k}} q(\bm{k}) \right]^3$ is the winding number characterizing the ground state topology of the chiral symmetric topological insulator \cite{Schnyder}.

\subsection{The winding number in arbitrary odd spacial dimensions}
The winding number $N_{2n+1}$ of the chiral symmetric topological insulator (superconductor) in $2n+1$ spacial dimensions 
characterize the homotopy of the map from BZ $\ni \bm{k} \mapsto q(\bm{k})$ in the unitary group $U(m)$, 
where $q(\bm{k})$ is the off-diagonal part of $Q(\bm{k})$ on the basis such that $\Gamma$ is represented as $\Gamma = \begin{pmatrix}
1 & 0 \\
0 & -1
\end{pmatrix}$. 
$N_{2n+1}$ is given by  
\begin{equation}
\begin{split}
N_{2 n+1} = -\frac{1}{(2 \pi i)^{n+1} 2^n (2 n+1)!!} \int \mathrm{tr} \left[ q^{\dag}(\bm{k}) d q(\bm{k}) \right]^{2 n+1}. 
\end{split}
\label{sup0}
\end{equation}
Eq. (\ref{sup0}) is rewritten as 
\begin{equation}
\begin{split}
N_{2 n+1} 
&= \frac{1}{(2 \pi i)^{n+1} 2^{n+1} (2 n+1)!!} \int \mathrm{tr} \ \Gamma \left[ Q(\bm{k}) d Q(\bm{k}) \right]^{2 n+1} \\
&= \frac{(-1)^n}{(2 \pi i)^{n+1} 2^{n+1} (2 n+1)!!} \int \mathrm{tr} \ \Gamma Q(\bm{k}) \left[ d Q(\bm{k}) \right]^{2 n+1} . \\
\end{split}
\label{WN2}
\end{equation}
Here, we used $Q(\bm{k}) d Q(\bm{k}) Q(\bm{k}) = - d Q(\bm{k})$ from $Q^2(\bm{k})=1$.

\subsection{The derivation of Eq. (15) and some algebraic properties of projected chiral position operator}
In this section, we will derive Eq. (15). 
For this purpose, we, first, explain some important features of the projected chiral position operator
$\hat{X}^5_{\mu}$
which hold both in 1D and 3D systems.
$\hat{X}^5_{\mu}$ has some similarity with the projected position operator $\hat X_{\mu} = P \hat r_{\mu} P$
where $P = \sum_{n \in \mathrm{occupied}} \sum_{\bm{k} \in \mathcal{BZ}} \ket{\phi_{n\bm{k}}} \bra{\phi_{n\bm{k}}}$ is a projector on the occupied bands. 
It is useful for the following argument to summarize some basic properties of $\hat X_{\mu}$ here.
On the basis of Bloch states $\ket{\phi_{n\bm{k}}}$, $\hat X_{\mu}$ is represented as 
$\Braket{\phi_{n\bm{k}} |\hat X_{\mu} | \phi_{m\bm{k}'}} = \left( i \delta_{nm} \frac{\partial}{\partial k_{\mu}} + \mathcal{A}_{nm,\mu}(\bm{k}) \right) \delta_{\bm{k},\bm{k}'}$, 
where $\mathcal{A}_{nm,\mu}(\bm{k}) = i \Braket{n_{\bm{k}} | \frac{\partial m_{\bm{k}}}{\partial k_{\mu}} }$ with $\ket{n_{\bm{k}}} = e^{-i \bm{k} \cdot \bm{r}} \ket{\phi_{n\bm{k}}}$ is the Berry connection. 
Then the non-commutativity of the projected position operator $\hat X_{\mu}$ yields 
$ \Braket{ \phi_{n\bm{k}} | \left[ \hat X_{\mu}, \hat X_{\nu} \right] | \phi_{m\bm{k}'}} = i \mathcal{F}_{nm,\mu \nu}(\bm{k}) \delta_{\bm{k},\bm{k}'} $, 
where $\mathcal{F}_{\mu \nu}(\bm{k}) = \partial_{\mu} \mathcal{A}_{\nu} - \partial_{\nu} \mathcal{A}_{\mu} + \left[ \mathcal{A}_{\mu} , \mathcal{A}_{\nu} \right] $ is the non-Abelian Berry curvature
arising in multi-band systems.  
Under a gauge transformation of the occupied Bloch states $\Psi(\bm{k}) = \left\{ \ket{n_{\bm{k}}}, \ket{m_{\bm{k}}}, \cdots \right\}_{n,m,\cdots \in occ} \mapsto \Psi(\bm{k}) U(\bm{k})$ with a unitary matrix $U(\bm{k})$, 
the Berry curvature $\mathcal{F}$ is transformed as $\mathcal{F} \mapsto U^{\dag} \mathcal{F} U$, hence a nontrivial gauge dependence do not exist. 

We now apply a similar argument to the projected chiral position operator defined by
\begin{equation}
\begin{split}
\hat X^5_{\mu} := P \Gamma \hat r_{\mu} P. 
\end{split}
\end{equation}
The representation of $\hat X^5_{\mu}$ on the basis of Bloch states is given by
\begin{equation}
\begin{split}
\Braket{\phi_{n\bm{k}} | \hat X^5_{\mu} | \phi_{m\bm{k}'}} 
&= \Braket{n_{\bm{k}} | e^{-i \bm{k} \cdot \hat{\bm{r}}} \Gamma \left( -i \frac{\partial}{\partial k'_{\mu}} e^{i \bm{k}' \cdot \hat{\bm{r}}} + i e^{i \bm{k}' \cdot \hat{\bm{r}}} \frac{\partial}{\partial k'_{\mu}} \right) | m_{\bm{k}'}}  \\
&= i \Braket{n_{\bm{k}}|\Gamma|\frac{\partial m_{\bm{k}}}{\partial k_{\mu}}} \delta_{\bm{k},\bm{k}'} . \\
\end{split}
\end{equation}
Note that $\Braket{\phi_{n\bm{k}} |\Gamma| \phi_{m\bm{k}'}}$ vanishes when
both $|\phi_{n \bm{k}}\rangle$ and $|\phi_{m \bm{k}'}\rangle$ are occupied states,
because of the chiral symmetry. 
We denote $\tilde X^5_{nm,\mu}(\bm{k}) := i \Braket{n_{\bm{k}}|\Gamma|\frac{\partial m_{\bm{k}}}{\partial k_{\mu}}}$ 
or $\tilde X^5_{\mu}(\bm{k}) := i \Psi^{\dag}(\bm{k}) \Gamma \partial_{\mu} \Psi(\bm{k})$ in matrix representation,  
Hereafter, $\tilde X^5_{\mu}(\bm{k}) $ is referred to as the projected chiral position.
Under a gauge transformation $\Psi(\bm{k}) \mapsto \Psi(\bm{k}) U(\bm{k})$, $\tilde X^5_{\mu}(\bm{k})$ transforms as 
\begin{equation}
\begin{split}
\tilde X^5_{\mu}(\bm{k}) 
&\mapsto i U^{\dag}(\bm{k}) \Psi^{\dag}(\bm{k}) \Gamma \partial_{\mu} \left\{ \Psi(\bm{k}) U(\bm{k}) \right\} \\
&= i U^{\dag}(\bm{k}) \Psi^{\dag}(\bm{k}) \Gamma \partial_{\mu} \Psi(\bm{k}) U(\bm{k}) + i U^{\dag}(\bm{k}) \Psi^{\dag}(\bm{k}) \Gamma \Psi(\bm{k}) \partial_{\mu} U(\bm{k}) \\
&= U^{\dag}(\bm{k}) \tilde X^5_{\mu}(\bm{k}) U(\bm{k}). 
\end{split}
\end{equation}
Thus, 
Here, we used $\Braket{n_{\bm{k}} | \Gamma | m_{\bm{k}}} = 0$ for the $n,m$ occupied bands. 
Hence, in the chiral symmetric systems, the projected chiral position $\tilde X^5_{\mu}(\bm{k})$ is gauge invariant in the same way as the Berry curvature $\mathcal{F}_{\mu\nu}(\bm{k})$, in sharp contrast to
 the projected position which is gauge-dependent.
Furthermore, 
the commutator of the projected chiral position yields the Berry curvature,
\begin{equation}
\begin{split}
\left[ \tilde X^5_{\mu}, \tilde X^5_{\nu} \right] 
&= - \Psi^{\dag} \Gamma \partial_{\mu} \Psi \Psi^{\dag} \Gamma \partial_{\nu} \Psi - (\mu \leftrightarrow \nu) \\
&= \partial_{\mu} \Psi^{\dag} \Gamma \Psi \Psi^{\dag} \Gamma \partial_{\nu} \Psi - (\mu \leftrightarrow \nu) \\
&= \partial_{\mu} \Psi^{\dag} \left( 1 - \Psi \Psi^{\dag} \right) \partial_{\nu} \Psi - (\mu \leftrightarrow \nu) \\
&= -i \mathcal{F}_{\mu \nu}. 
\end{split}
\label{x5x5}
\end{equation}
Here we used 
\begin{equation}
\begin{split}
1 = \sum_{n}\ket{n}\bra{n} = \sum_{n \in occpied}\ket{n}\bra{n} + \sum_{\bar n \in unoccupied}\ket{\bar n}\bra{\bar n} = \sum_{n \in occupied} \Big[ \ket{n}\bra{n} + \Gamma \ket{n}\bra{n} \Gamma \Big] = \Psi \Psi^{\dag} + \Gamma \Psi \Psi^{\dag} \Gamma .
\end{split}
\end{equation}

We also remark that the winding number is written as the integral of the $(2n+1)$-bracket of the projected chiral position over the Brillouin zone : 
\begin{equation}
\begin{split}
N_{2n+1}
&= \frac{-i^n}{\pi^{n+1} (2n+1) !!} \int_{BZ} \mathrm{tr} \left[ \tilde X^5_{1} , \tilde X^5_{2} , \cdots , \tilde X^5_{2n+1}  \right] \\
&= \frac{-i^n \epsilon_{\mu_1 \mu_2 \cdots \mu_{2n+1}}}{\pi^{n+1} (2n+1) !!} \int_{\mathrm{BZ}} d^{2n+1} k \ \mathrm{tr} \left[ \tilde X^5_{\mu_1}, \tilde X^5_{\mu_2}, \cdots ,\tilde X^5_{\mu_{2n+1}} \right], 
\end{split}
\label{WBRA}
\end{equation}
where $\left[ X_{1}, X_{2}, \cdots, X_{2n+1} \right] = \epsilon_{\mu_1 \mu_2 \cdots \mu_{2n+1}} X_{\mu_1} X_{\mu_2} \cdots X_{\mu_{2n+1}}$ is the $(2n+1)$-bracket, and $\mu_i$ run over $1, 2, \cdots , 2n+1$. 
Note that when $n=1$ (i.e. $N_3$), $[X_1, X_2, X_3]$ is the Nambu bracket \cite{Nambu}.
This expression of the winding number is analogous to the relation between the Chern number and non-commutativity of the projected position operator. 
Eq.(\ref{WBRA}) directly follows from (\ref{WN2}). 
The matrix element of $d Q = d \left( 1 - 2 \sum_{n \in occupied} \ket{n}\bra{n} \right) = -2 \sum_{n \in occupied} \left( \ket{dn}\bra{n} + \ket{n}\bra{dn} \right) $ 
on the basis of the occupied states $\ket{n}$ and unoccupied states $\ket{\bar n} = \Gamma \ket{n}$ are 
\begin{equation}
\begin{split}
&\Braket{n | d Q | n'} = 0 ,\\
&\Braket{\bar n | d Q | \bar n'} = 0 ,\\
&\Braket{n | d Q | \bar n'} = 2 \Braket{n | \Gamma | d n'} ,\\
&\Braket{\bar n | d Q | n'} = -2 \Braket{n | \Gamma | d n'} .\\
\end{split}
\end{equation}
Hence $dQ$ is written as the matrix form in the space spanned by the occupied and unoccupied states, 
\begin{equation}
\begin{split}
d Q &\rightarrow  
\begin{pmatrix}
\Braket{n | d Q | n'} & \Braket{n | d Q | \bar n'}\\
\Braket{\bar n | d Q | n'} & \Braket{\bar n | d Q | \bar n'}
\end{pmatrix}
= \begin{pmatrix}
0 & 2 \Braket{n | \Gamma | d n'} \\
-2 \Braket{n | \Gamma | d n'} & 0
\end{pmatrix} \\
&= - 2 i \tau_2 \Braket{n | \Gamma | d n'} = -2 \tau_2 \tilde X^5_{nn',\mu} \ d k_{\mu}
\end{split}
\label{OU1}
\end{equation}
where $\bm{\tau} = (\tau_1, \tau_2, \tau_3)$ is the Pauli matrices in the occupied-unoccupied space. 
In the same way, $\Gamma$ and $Q$ is written as 
\begin{equation}
\begin{split}
\Gamma \rightarrow 
\begin{pmatrix}
\Braket{n | \Gamma | n'} & \Braket{n | \Gamma | \bar n'}\\
\Braket{\bar n | \Gamma | n'} & \Braket{\bar n | \Gamma | \bar n'}
\end{pmatrix}
= \begin{pmatrix}
0 & \delta_{nn'} \\
\delta_{nn'} & 0
\end{pmatrix} 
= \tau_1 \delta_{nn'}, 
\end{split}
\label{OU2}
\end{equation}
\begin{equation}
\begin{split}
Q \rightarrow 
\begin{pmatrix}
\Braket{n | Q | n'} & \Braket{n | Q | \bar n'}\\
\Braket{\bar n | Q | n'} & \Braket{\bar n | Q | \bar n'}
\end{pmatrix}
= \begin{pmatrix}
-\delta_{nn'} & 0 \\
0 & \delta_{nn'}
\end{pmatrix} 
= -\tau_3 \delta_{nn'}. 
\end{split}
\label{OU3}
\end{equation}
By using (\ref{OU1}), (\ref{OU2}) and (\ref{OU3}), we can write 
the winding number (\ref{WN2}) as 
\begin{equation}
\begin{split}
N_{2 n+1} 
&= \frac{(-1)^n}{(2 \pi i)^{n+1} 2^{n+1} (2 n+1)!!} \int_{BZ} \mathrm{tr} \ \Gamma Q(\bm{k}) \left[ d Q(\bm{k}) \right]^{2 n+1} \\
&= \frac{(-1)^n}{(2 \pi i)^{n+1} 2^{n+1} (2 n+1)!!} \int_{BZ} \mathrm{tr} \left[ \tau_1 (-\tau_3) \left( -2 \tau_2 \tilde X^5_{\mu_1} \ d k_{\mu_2} \right) \left( -2 \tau_2 \tilde X^5_{\mu_2} \ d k_{\mu_2} \right) \cdots \left( -2 \tau_2 \tilde X^5_{\mu_{2n+1}} \ d k_{\mu_{2n+1}} \right) \right] \\
&= \frac{(-1)^{n+1}}{(\pi i)^{n} (2 n+1)!!} \int_{BZ} \mathrm{tr} \left[ \tilde X^5_{\mu_1} \tilde X^5_{\mu_2} \cdots \tilde X^5_{\mu_{2n+1}} \right] d k_{\mu_1} d k_{\mu_2} \cdots d k_{\mu_{2n+1}} \\
&= \frac{(-1)^{n+1} \epsilon_{\mu_1 \mu_2 \cdots \mu_{2n+1}}}{(\pi i)^{n} (2 n+1)!!} \int_{BZ}  d^{2n+1} \bm{k} \ \mathrm{tr}\left[ \tilde X^5_{\mu_1} \tilde X^5_{\mu_2} \cdots \tilde X^5_{\mu_{2n+1}} \right], \\
\end{split}
\end{equation}
Thus, we arrive at (\ref{WBRA}). 

In the case of the one dimensional systems, the winding number 
\begin{equation}
\begin{split}
N_1 = -\frac{1}{\pi} \int_{BZ} \mathrm{tr} \ \tilde X^5(\bm{k})
\end{split}
\end{equation}
is directly related to the chiral polarization defined by (14).  
For the Wannier function localized at the site $R$, (14) is 
\begin{equation}
\begin{split}
P^5 
&= \frac{e}{V_c N_c} \sum_{n \in occupied} \sum_{k,k'} \Braket{\phi_{n k}| e^{i k R} \Gamma \hat x e^{-i k' R} | \phi_{n k'}} \\
&= \frac{e}{V_c N_c} \sum_{n \in occupied} \sum_{k,k'} e^{i (k-k') R} \ \mathrm{tr} \ \tilde X^5(k) \delta_{k,k'} \\
&= \frac{e}{2 \pi} \int_{BZ} d k \ \mathrm{tr} \ \tilde X^5(k) \\
&= -\frac{N_1 e}{2}, 
\end{split}
\end{equation}
where $N_c$ is the number of unit cell. 
Thus we obtain Eq.(15). 

To close this section, we would like to comment on an implication of (\ref{WBRA}) to the 3D case. In this case,
from Eq.(\ref{WBRA}), the winding number $N_3$ is expressed in terms of the Nambu bracket.
This implies that $N_3$ may be related to a physical quantity described by the Nambu dynamics \cite{Nambu}.
However, we do not know any examples in condensed matter systems which are 
described by the Nambu mechanics, and also quantum version of the Nambu mechanics is not well understood.
We have not yet succeeded to obtain any insight from the study in this direction.
Thus, we consider a different approach to relate the 3D winding number $N_3$ to a physical quantity, as explained
in the main text and the following section.

\subsection{The derivation of Eq. (16)}
In this section, we derive Eq.(16) on the assumption that the occupied and unoccupied Wannier states satisfy the chiral symmetry 
$\Ket{w_{\bar n\bm{R}}} = \Gamma \Ket{w_{n\bm{R}}}$ ($\bar n \in \text{unoccupied}, n \in \text{occupied}$ ).
This is equivalent to the gauge fixing condition $\ket{\phi_{\bar n \bm{k}}} = \Gamma \ket{\phi_{n\bm{k}}}$ for unoccupied and occupied Bloch states.
Our derivation of Eq.(16) is based on the perturbation formalism developed by Kita-Arai \cite{Kita} for the Wannier function under the magnetic field. 
Generally, to construct the exponentially localized Wannier function, we need the absence of gauge obstruction of the Bloch wave function, 
i.e., vanishing of Chern number $C_{ij}/(2 \pi i) = \int_{BZ} d^3k/(2 \pi)^3 \mathrm{tr} \mathcal{F}_{ij} = 0$. 
In chiral symmetric systems, the Chern numbers $C_{ij}$ are zeroes, 
and hence the exponentially localized Wannier functions are always well defined. 

The chiral charge polarization in the case with no magnetic field introduced in the main text is 
\begin{equation}
\begin{split}
\bm{P}^5_{( \bm{B}=\bm{0} )} = \frac{e}{V_c} \sum_{n\in occ}\Braket{w_{n\bm{R}}|\Gamma \hat{\bm{r}}|w_{n\bm{R}}}, 
\end{split}
\end{equation}
where $\Ket{w_{n\bm{R}}}$ is the Wannier function localized at a site $\bm{R}$ constructed from the $n$-th occupied band, 
\begin{equation}
\begin{split}
\Ket{w_{n\bm{R}}} = \frac{1}{\sqrt{N_c}} \sum_{\bm{k}} e^{-i \bm{k} \cdot \bm{R}} \Ket{\phi_{n\bm{k}}},
\end{split}
\end{equation}
where $N_c$ is the number of unit cells. 
Now, we will calculate the first order perturbative corrections to $\bm{P}^5$ with respect to an applied uniform magnetic field $\bm{B}$. 
First, we introduce modified Wannier states $\ket{w'_{n\bm{R}}}$ defined by \cite{Luttinger}
\begin{equation}
\begin{split}
w'_{n\bm{R}}(\bm{r}) = e^{i I_{\bm{r}\bm{R}}} w_{n\bm{R}}(\bm{r}), 
\end{split}
\end{equation}
where $I_{\bm{r}\bm{R}}$ is the Peierls phase  
\begin{equation}
\begin{split}
I_{\bm{r}\bm{R}} = \frac{e}{\hbar c} \int_{\bm{R}}^{\bm{r}} d \bm{r}' \cdot \bm{A}(\bm{r}') 
\end{split}
\end{equation}
with $d\bm{r}'$ the straight line path from $\bm{R}$ to $\bm{r}$. 
The modified Wannier states $\ket{w'_{n\bm{R}}}$ form a complete set, though they are not orthonormal for the case with 
a finite magnetic field. 
To orthonormalize them,  we need to include corrections from other sites and other bands to $\ket{w'_{n\bm{R}}}$.
Then, the orthonormal modified Wannier function is expressed as,
\begin{equation}
\begin{split}
\ket{\varphi_{n\bm{R}}} = \sum_{n'\bm{R}} \ket{w'_{n'\bm{R}'}} S_{n'\bm{R}',n\bm{R}},
\end{split}
\end{equation}
for which $\braket{\varphi_{n\bm{R}}|\varphi_{n'\bm{R}'}} = \delta_{nn'} \delta_{\bm{R}\bm{R}'}$ is satisfied. 
Thus, the chiral charge polarization under an applied magnetic field is given by 
\begin{equation}
\begin{split}
\bm{P}^5 = \frac{e}{V_c} \sum_{n\in occ}\Braket{\varphi_{n\bm{R}}|\Gamma \hat{\bm{r}}|\varphi_{n\bm{R}}}. 
\end{split}
\end{equation}
From Ref. \cite{Kita}, in the cases of a uniform magnetic field, $S_{n'\bm{R}',n\bm{R}}$ is expanded up to the first order in $\bm{B}$:
\begin{equation}
\begin{split}
S_{n'\bm{R}',n\bm{R}} = \delta_{nn'} \delta_{\bm{R}\bm{R}'} - \frac{i e}{4 \hbar c} B_i \epsilon_{ijl} \frac{e^{i I_{\bm{R}'\bm{R}}}}{N_c} \sum_{\bm{k}} e^{i \bm{k} \cdot (\bm{R}'-\bm{R})} \braket{\partial_j n'_{\bm{k}}|\partial_l n_{\bm{k}}}, 
\end{split}
\label{sup1}
\end{equation}
where $\ket{n_{\bm{k}}} = e^{-i \bm{k} \cdot \hat{\bm{r}}} \ket{\phi_{n \bm{k}}}$. 
We denote $\ket{\varphi_{n\bm{R}}} = \ket{w'_{n\bm{R}}} + \ket{\delta w'_{n\bm{R}}}$.  Then, the correction term of the chiral polarization is 
\begin{equation}
\begin{split}
\delta \bm{P}^5 = \frac{e}{V_c} \sum_{n\in occ} \left[ \Braket{w'_{n\bm{R}}|\Gamma \hat{\bm{r}}|w'_{n\bm{R}}} - \Braket{w_{n\bm{R}}|\Gamma \hat{\bm{r}}|w_{n\bm{R}}} \right] +\frac{e}{V_c} \sum_{n\in occ} \left[ \Braket{w'_{n\bm{R}}|\Gamma \hat{\bm{r}}|\delta w'_{n\bm{R}}} + c.c. \right]. 
\end{split}
\end{equation} 
The first term vanishes since the Peierls phases of $\ket{w'_{n\bm{R}}}$ are canceled out in $\Braket{w'_{n\bm{R}}|\Gamma \hat{\bm{r}}|w'_{n\bm{R}}}$. 
The second term is recast into the following form with the use of (\ref{sup1}),
\begin{equation}
\begin{split}
\Braket{w'_{n\bm{R}}|\Gamma \hat{\bm{r}}|\delta w'_{n\bm{R}}}
&= - \frac{i e}{4 \hbar c} B_i \epsilon_{ijl} \sum_{n'\bm{R}'} \Braket{w'_{n\bm{R}}|\Gamma \hat{\bm{r}}|w'_{n'\bm{R}'}} \frac{e^{i I_{\bm{R}'\bm{R}}}}{N_c} \sum_{\bm{k}} e^{i \bm{k} \cdot (\bm{R}'-\bm{R})} \braket{\partial_j n'_{\bm{k}}|\partial_l n_{\bm{k}}} \\
&= - \frac{i e}{4 \hbar c} B_i \epsilon_{ijl} \sum_{n'\bm{R}'} \Braket{w_{n\bm{R}}|\Gamma \hat{\bm{r}}|w_{n'\bm{R}'}} \frac{1}{N_c} \sum_{\bm{k}} e^{i \bm{k} \cdot (\bm{R}'-\bm{R})} \braket{\partial_j n'_{\bm{k}}|\partial_l n_{\bm{k}}}.  \\
\end{split}
\label{eq:41}
\end{equation}
Here, we omitted the Peierls phases since they are higher order corrections. 
On the other hand, the factor $\Braket{w_{n\bm{R}}|\Gamma \hat{\bm{r}}|w_{n'\bm{R}'}}$ is expressed as,
\begin{equation}
\begin{split}
\Braket{w_{n\bm{R}}|\Gamma \hat{\bm{r}}|w_{n'\bm{R}'}}
&= \frac{1}{N_c} \sum_{\bm{k},\bm{k}'} e^{i \bm{k} \cdot \bm{R}} \Braket{\phi_{n\bm{k}} |\Gamma \hat{\bm{r}} | \phi_{n'\bm{k}'}} e^{-i \bm{k}' \cdot \bm{R}'} \\
&= \frac{1}{N_c} \sum_{\bm{k},\bm{k}'} e^{i \bm{k} \cdot \bm{R}} \left\{ -i \frac{\partial}{\partial \bm{k}'} \left( \braket{n_{\bm{k}}|\Gamma|n'_{\bm{k}}} \delta_{\bm{k},\bm{k}'} \right) + i \braket{n_{\bm{k}}|\Gamma|\bm{\nabla} n'_{\bm{k}}} \delta_{\bm{k},\bm{k}'} \right\} e^{-i \bm{k}' \cdot \bm{R}'} \\
&= \bm{R} \delta_{\bar n n'} \delta_{\bm{R},\bm{R}'} + \frac{1}{N_c} \sum_{\bm{k}} e^{i \bm{k} \cdot (\bm{R}-\bm{R}')} i \braket{n_{\bm{k}}|\Gamma| \bm{\nabla} n'_{\bm{k}}},  \\
\end{split}
\label{eq:42}
\end{equation}
where $\bar n$ is a label for an unoccupied state and we fixed the gauge of $\ket{\phi_{\bar n}}$ satisfying $\ket{\phi_{\bar n}} = \Gamma \ket{\phi_n}$ with $\ket{\phi_n}$ an occupied state.  
From (\ref{eq:41}) and (\ref{eq:42}), we have, 
\begin{equation}
\begin{split}
\Braket{w'_{n\bm{R}}|\Gamma \hat{\bm{r}}|\delta w'_{n\bm{R}}} 
&= - \frac{i e}{4 \hbar c} B_i \epsilon_{ijl} \frac{1}{N_c} \sum_{\bm{k}} \left[ \bm{R} \braket{\partial_j n'_{\bm{k}}|\Gamma|\partial_l n_{\bm{k}}} + \sum_{n'} i \braket{n_{\bm{k}}|\Gamma|\bm{\nabla} n'_{\bm{k}}} \braket{\partial_j n'_{\bm{k}}|\partial_l n_{\bm{k}}} \right].  \\
\end{split}
\end{equation}
The first term vanishes because it is the total derivative : $\epsilon_{ijl} \braket{\partial_j n_{\bm{k}}|\Gamma|\partial_l n_{\bm{k}}} = \epsilon_{ijl} \partial_j \left\{ \braket{n_{\bm{k}}|\Gamma|\partial_l n_{\bm{k}}} \right\}$. 
Hence we obtain, 
\begin{equation}
\begin{split}
\delta \bm{P}^5 
&= \frac{e^2}{4 \hbar c} B_i \epsilon_{ijl} \frac{1}{N_c V_c} \sum_{n \in occ} \sum_{n'\bm{k}} \braket{n_{\bm{k}}|\Gamma|\bm{\nabla} n'_{\bm{k}}} \braket{\partial_j n'_{\bm{k}}|\partial_l n_{\bm{k}}} + c.c. \\
&= \frac{e^2 }{2 \hbar c } B_i \epsilon_{ijl} \int \frac{d^3 \bm{k}}{(2 \pi)^3} \mathrm{tr} \left[ \Psi^{\dag} \Gamma \bm{\nabla} \Psi \partial_j \Psi^{\dag} \partial_l \Psi + \Psi^{\dag} \bm{\nabla} \Psi \partial_j \Psi^{\dag} \Gamma \partial_l \Psi  \right], 
\end{split}
\label{eq:44}
\end{equation}
where $\Psi = \left\{ \ket{n_1}, \ket{n_2}, \cdots \right\}_{n_i \in occ}$ and we omit the parameter $\bm{k}$.

From this expression, we find that $\alpha_{ij} = \frac{\delta P_i^5 }{ \delta B_j}$ is proportional to $\delta_{ij}$. 
For instance, we consider an off-diagonal component, 
\begin{equation}
\begin{split}
\alpha_{xy} 
&= \frac{e^2}{2 \hbar c} \epsilon_{yjl} \int \frac{d^3 \bm{k}}{(2 \pi)^3} \mathrm{tr} \left[ \Psi^{\dag} \Gamma \partial_x \Psi \partial_j \Psi^{\dag} \partial_l \Psi + \Psi^{\dag} \partial_x \Psi \partial_j \Psi^{\dag} \Gamma \partial_l \Psi \right] \\
&= \frac{e^2}{2 \hbar c} \int \frac{d^3 \bm{k}}{(2 \pi)^3} \mathrm{tr} \left[\Psi^{\dag} \Gamma \partial_x \Psi \partial_z \Psi^{\dag} \partial_x \Psi - \Psi^{\dag} \Gamma \partial_x \Psi \partial_x \Psi^{\dag} \partial_z \Psi + \Psi^{\dag} \partial_x \Psi \partial_z \Psi^{\dag} \Gamma \partial_x \Psi - \Psi^{\dag} \partial_x \Psi \partial_x \Psi^{\dag} \Gamma \partial_z \Psi \right] . \\
\end{split}
\label{sup12}
\end{equation}
We see that on the interior of the trace in Eq.(\ref{sup12}), 
\begin{equation}
\begin{split}
(1st) + (4th) 
&= \mathrm{tr} \left[ \Psi^{\dag} \Gamma \partial_x \Psi \partial_z \Psi^{\dag} \partial_x \Psi - \Psi^{\dag} \partial_x \Psi \partial_x \Psi^{\dag} \Gamma \partial_z \Psi \right] \\
&= \mathrm{tr} \left[ -\partial_x \Psi \partial_x \Psi^{\dag} \Gamma \left( \Psi \partial_z \Psi^{\dag} + \partial_z \Psi \Psi^{\dag} \right) \right] \\
&= \mathrm{tr} \left[ -\partial_x \Psi \partial_x \Psi^{\dag} \Gamma \partial_z P \right] , \\
\end{split}
\end{equation}
and
\begin{equation}
\begin{split}
(2nd) + (3rd) 
&= \mathrm{tr} \left[ -\Psi^{\dag} \Gamma \partial_x \Psi \partial_x \Psi^{\dag} \partial_z \Psi + \Psi^{\dag} \partial_x \Psi \partial_z \Psi^{\dag} \Gamma \partial_x \Psi \right] \\
&= \mathrm{tr} \left[ -\partial_x \Psi \partial_x \Psi^{\dag} \left( \partial_z \Psi \Psi^{\dag} + \Psi \partial_z \Psi^{\dag} \right) \Gamma \right] \\
&= \mathrm{tr} \left[ -\partial_x \Psi \partial_x \Psi^{\dag} \partial_z P \Gamma \right]. \\
\end{split}
\end{equation}
Thus, we can show 
\begin{equation}
\begin{split}
\alpha_{xy} = \frac{e^2}{2 \hbar c} \int \frac{d^3 \bm{k}}{(2 \pi)^3} \mathrm{tr} \left[ -\partial_x \Psi \partial_x \Psi^{\dag} \left( \Gamma \partial_z P + \partial_z P \Gamma \right) \right] = 0 , \\
\end{split}
\end{equation}
where, $P = \sum_{n \in occ} \ket{n}\bra{n}$ is a projector to occupied bands, and we have used $d \Psi \Psi^{\dag} + \Psi d \Psi^{\dag} = d P$, and $\Gamma \partial_z P + \partial_z P \Gamma = \partial_z \Gamma = 0$. 
Hence, 
\begin{equation}
\begin{split}
\alpha_{ij} &= \delta_{ij} \frac{e^2}{48 \pi^3 \hbar c} \int \mathrm{tr} \left[ \Psi^{\dag} \Gamma d \Psi d \Psi^{\dag} d \Psi + \Psi^{\dag} d \Psi d \Psi^{\dag} \Gamma d \Psi \right] \\
&= \delta_{ij} \frac{e^2}{24 \pi^3 \hbar c} \int \mathrm{tr} \left[ \Psi^{\dag} \Gamma d \Psi d \Psi^{\dag} d \Psi \right] , \\
\end{split}
\end{equation}
The integrand of this expression is gauge-invariant under gauge transformation $\Psi \rightarrow \Psi U$ with a unitary matrix $U$  except for total derivative.
This is seen as follows. 
\begin{equation}
\begin{split}
&\mathrm{tr} \left[ \Psi^{\dag} \Gamma d \Psi d \Psi^{\dag} d \Psi \right] \\
&\mapsto \mathrm{tr} \left[ U^{\dag} \Psi^{\dag} \Gamma d ( \Psi U ) d (U^{\dag} \Psi^{\dag} ) d ( \Psi U ) \right] \\
&= \mathrm{tr} \left[ \Psi^{\dag} \Gamma d \Psi \left( d\Psi^{\dag} d \Psi + U d U^{\dag} \Psi^{\dag} d \Psi + \Psi^{\dag} d \Psi U d U^{\dag} -U d U^{\dag} U d U^{\dag} \right) \right] \\
&= \mathrm{tr} \left[ \Psi^{\dag} \Gamma d \Psi d\Psi^{\dag} d \Psi \right] + \mathrm{tr} \left[ \Psi^{\dag} d \Psi \Psi^{\dag} \Gamma d \Psi U d U^{\dag} + \Psi^{\dag} \Gamma d \Psi \Psi^{\dag} d \Psi U d U^{\dag} + \Psi^{\dag} \Gamma d \Psi d U d U^{\dag} \right] \\
&= \mathrm{tr} \left[ \Psi^{\dag} \Gamma d \Psi d\Psi^{\dag} d \Psi \right] + \mathrm{tr} \left[ - d \Psi^{\dag} \left( \Psi \Psi^{\dag} \Gamma + \Gamma \Psi \Psi^{\dag} \right) d \Psi U d U^{\dag} + \Psi^{\dag} \Gamma d \Psi d U d U^{\dag} \right] \\
&= \mathrm{tr} \left[ \Psi^{\dag} \Gamma d \Psi d\Psi^{\dag} d \Psi \right] + \mathrm{tr} \left[ - d \Psi^{\dag} \Gamma d \Psi U d U^{\dag} + \Psi^{\dag} \Gamma d \Psi d U d U^{\dag} \right] \\
&= \mathrm{tr} \left[ \Psi^{\dag} \Gamma d \Psi d\Psi^{\dag} d \Psi \right] - d \ \mathrm{tr} \left[ \Psi^{\dag} \Gamma d \Psi U d U^{\dag} \right] , \\
\end{split}
\end{equation}
where, we have used $\Psi \Psi^{\dag} \Gamma + \Gamma \Psi \Psi^{\dag} = \sum_{n \in occ} \left( \ket{n}\bra{n} \Gamma + \Gamma \ket{n}\bra{n} \right) = \Gamma$. 
Hence the integral over the Brillouin zone is gauge-invariant. 
Now, we choose the basis, 
\begin{equation}
\begin{split}
\Psi = \frac{1}{\sqrt{2}} \begin{pmatrix}
q \\
-1
\end{pmatrix}, 
\end{split}
\end{equation}
where, $q$ is the off diagonal part of Q-function $Q = \begin{pmatrix}
0 & q \\
q^{\dag} & 0
\end{pmatrix}$. 
Then 
\begin{equation}
\begin{split}
\alpha^i_j &= -\delta^i_j \frac{e^2}{96 \pi^3 \hbar} \int \mathrm{tr} (q^{\dag} d q)^3 \\
&= -\delta^i_j \frac{e^2}{4 \pi \hbar} N_3, \\
\end{split}
\end{equation}
where $N_3 = \frac{1}{24 \pi^3} \int \mathrm{tr} (q^{\dag} d q)^3$ is the winding number. 
Thus we arrive at Eq. (16). 

Note that the magnetic induced chiral polarization represented by the first line of (\ref{eq:44}) generally depends on the gauge of the unoccupied Bloch states $\ket{\phi_{\bar n \bm{k}}}$,  
so the gauge fixing condition $\ket{\phi_{\bar n\bm{k}}} = \Gamma \ket{\phi_{n\bm{k}}}$ is crucial for our results.

\end{document}